# FOS: A fully integrated open-source program for Fast Optical Spectrum calculations of nanoparticle media

Daniel Carne, Joseph Peoples, Ziqi Guo, Dudong Feng, Zherui Han, Xiaojie Liu, Xiulin Ruan[,*]

School of Mechanical Engineering and Birck Nanotechnology Center, Purdue University, West Lafayette, IN 47907, USA

[*] Corresponding Author: ruan@purdue.edu

**Abstract**

FOS, which means light in Greek, is an open-source program for Fast Optical Spectrum calculations of nanoparticle media. This program takes the material properties and a description of the system as input, and outputs the spectral response including the reflectance, absorptance, and transmittance. Previous open-source codes often include only one portion of what is needed to calculate the spectral response of a nanoparticulate medium, such as Mie theory or a Monte Carlo method. FOS is designed to provide a convenient fully integrated format to remove the barrier as well as providing a significantly accelerated implementation with compiled Python code, parallel processing, and pre-trained machine learning predictions. This program can accelerate optimization and high throughput design of optical properties of nanoparticle or nanocomposite media, such as radiative cooling paint and solar heating liquids, allowing for the discovery of new materials and designs. FOS also enables convenient modeling of lunar dust coatings, combustion particulates, and many other particulate systems. In this paper we discuss the methodology used in FOS, features of the program, and provide four case studies.

**Program Summary**

*Program Title:* FOS: Fast Optical Spectrum calculations of nanoparticle media

*CPC Library link to program files:* (to be added by Technical Editor)

*Developer's repository link:* https://github.com/FastOpticalSpectrum/FOS

*Licensing provisions:* GNU General Public License version 3

*Programming language:* Python 3.10

*Nature of problem:* Calculation of scattering and absorption properties, and the spectral response of nanoparticle media.

*Solution method:* Calculation of scattering and absorption properties is done through Mie theory or can be pre-calculated by the user. Calculation of the spectral response is done through either Monte Carlo simulations or a Machine Learning method.

*Additional comments including restrictions and unusual features:* The executable program is built for Windows computers. The Python code can be used for other operating systems such as Mac, Linux, or Unix.

*Preprint submitted to Computer Physics Communications.*

**Introduction**

Simulation and modeling of nanoparticle media's spectral optical response play a key role in the research and design of many applications. Recently, modeling of radiative cooling paint has allowed for the fast growth of the field including testing new materials, pigment shapes and sizes, and colored paint pigments [1], [2], [3], [4], [5], [6]. Simulation of nanoparticle media has also been used to model solar absorbers [7], phase change applications such as smart windows [8], [9], [10], and biomedical applications including sensing and imaging [11], [12]. However, calculation of nanoparticle media's spectral optical response including reflectance, absorptance, and transmittance can be a challenging task due to the complexity of the physics involved and the computational expense.

There are many methods to simulate the spectral response of a nanoparticle composite medium. One approach is to numerically solve Maxwell's equations on the nanoparticle geometry such as with Finite Volume (FV) or Finite-Difference Time-Domain (FDTD) methods [13]. However, these approaches are computationally expensive. One of the more common approaches, especially for coatings such as radiative cooling paint, is by first calculating the scattering and absorption properties from the material properties using either Mie theory or numerical methods [14], [15], [16], [17]. Subsequently, a variety of approaches can be used to solve the Radiative Transfer Equation (RTE) for the spectral response including deterministic and non-deterministic approaches [18]. Deterministic approaches like the Kubelka-Munk method [19], [20], [21] based on a two-flux model, the zonal method [22], [23], the Discrete Ordinates Method [24], [25], and the adding-doubling method [26], [27] are often efficient and reliable methods for solving the RTE. However, these methods often rely upon simplifications or approximations, such as the original Kubelka-Munk method assuming diffuse irradiation [20], and the original zonal method relying on isotropic scattering [18]. Alternatively, Monte Carlo simulations can be used as a non-deterministic method for solving the RTE without requiring simplifications or approximations [28], [29], [30]. Many different variants of Monte Carlo radiation transport simulations exist such as examples with variance reduction techniques [31], [32]. While Monte Carlo simulations are often slower than deterministic methods, there are also several key benefits. Primarily, the solution is not based on approximations or biased due to discretization, and the error estimation is easily quantified [33]. Additionally, Monte Carlo simulations are well suited for parallel computing due to individual photons not needing to communicate between each other [33]. Besides these methods, machine learning methods are also commonly used to accelerate solutions to various types of simulation methods [34], [35], [36], including specifically for solving the radiative transfer equation [37], [38], [39].

In our implementation, Mie theory is used to calculate the scattering and absorption properties of individual particles. Alternatively, a user can input their own scattering and absorption properties if they choose to calculate those separately, often in the case of particles of irregular shapes for which analytical Mie solutions are not available but numerical solutions are needed. Subsequently, the photon transport within the medium is modeled through Monte Carlo simulations or machine learning methods [40]. This provides the spectral response where important information can be calculated such as the coating color, solar reflectance, or sky window emittance. Using Mie theory and Monte Carlo together is a well-known process for calculating the spectral response of particulate media. Huang et al. used this process to investigate $TiO_2$ particle size for double-layer radiative cooling applications [6]. Peoples et al. furthered this work by showing multiple particle sizes of $TiO_2$ nanoparticles increased solar reflectance [1]. New materials for sub-ambient daytime radiative cooling have been discovered with the assistance of these simulation methods including $BaSO_4$, $CaCO_3$, and hBN [2], [4], [41]. Mie theory combined with Monte Carlo simulations is also aiding the search for efficient colored radiative cooling paints [5]. There are a few open-source Mie theory programs including miepython by Prahl [42], PyMieScatt by Sumlin et al. [43], and "MATLAB Functions or Mie Scattering and Absorption" by Mätzler [44]. There are also a few prominent open-source Monte Carlo radiation transport programs such as MCML by Wang et al. [45], OpenMC by Romano et al. [46], and MCmatlab by Marti et al. [47]. Many of these open-source codes often encompass only Mie theory or Monte Carlo photon transport, not both. In addition, implementations with significant acceleration, such as via machine learning, are desired.

To assist in removing these barriers and to provide open access to powerful simulation and machine learning-accelerated radiative transport tools, we present FOS. FOS, which means light in Greek, stands for Fast Optical Spectrum calculations for nanoparticle media. This open-source code includes accelerated Mie theory solutions to spherical and core-shell particles, can handle multi-layer media with multiple different particulate materials, and includes photon transport through parallel processing Monte Carlo simulations. Additionally, this code includes a pre-trained machine learning method which provides greatly accelerated photon transport predictions 1-3 orders of magnitude faster than Monte Carlo simulations [40]. These accelerated methods along with the all-in-one packaging for simulating the spectral response of nanoparticle media will allow for convenient access, and enhance optimization and high throughput design for applications such as thin films, radiative cooling paints, and spectrally selective materials. The paper includes the methodology used in FOS, information on how to use the program, and four examples of uses. Further information can be found in the GitHub repository

(https://github.com/FastOpticalSpectrum/FOS), and the input files from each example can be found in the supplementary information.

**Methodology**

As shown in Fig. 1, FOS is composed of five main elements, input, preprocessing, calculation of scattering properties, photon transport, and output. Each of these steps are described in detail below.

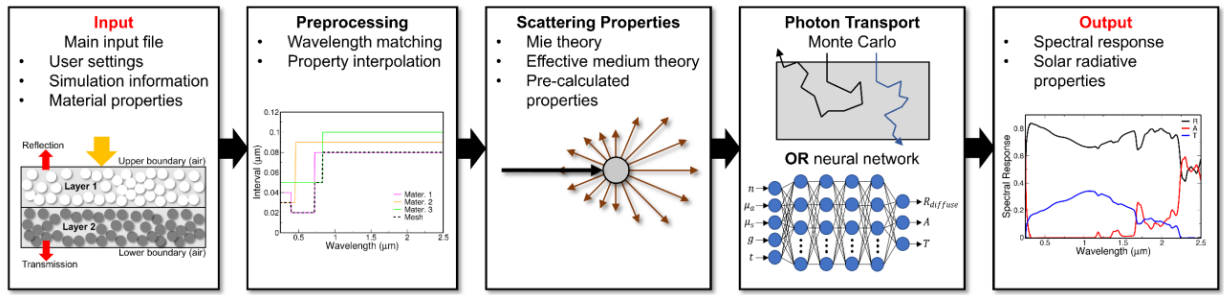

Fig. 1: Flowchart of FOS including the required inputs, preprocessing, calculation of optical properties, photon transport, and the output.

*Input*

FOS has one main input file consisting of two parts, a header and a body, with an entire example shown in section 1 of the supplementary information as well as in the GitHub repository. The header is used to import material files and set parameters that apply to every simulation, while the body includes each simulation with details on the materials and design. Throughout the entire input file, upper/lower case does not matter, and hashtags are used for comments. Below, an example of the header and body are shown along with a schematic showing this example in Fig. 2. In the first line of the header either "MC" or "NN" must be specified, referring to whether to use Monte Carlo (MC) simulations or Neural Network (NN) predictions for the photon transport. Next, the output file prefix is specified by "output: name". Here, each simulation would be saved as name1.txt, name2.txt, etc. Next, the particle and matrix material files are imported by "Particle 1: TiO2.txt" and "Matrix 1: air.txt". In this example, use of "Particle 1" in the input file will refer to the TiO2.txt material properties. Each particle material file can either contain the refractive index or the pre-calculated scattering properties. If entering the refractive index, the material file will consist of three columns including the wavelength, the refractive index and extinction coefficient at each wavelength. If entering the pre-calculated scattering properties, the material file will consist of four columns including the wavelength, the absorption coefficient, scattering coefficient, and asymmetry parameter at each wavelength. The units are based in µm. Examples of each type of material file are shown in section 2 of the supplementary information. Although not required, to integrate the spectral response of each simulation for calculating the total properties in the solar spectrum, include "solar: am15.txt" where the file titled am15.txt here includes two columns, the wavelength and the spectral solar power. If using Monte Carlo simulations, the number of photons per wavelength simulated is set by "photons: 30000". Finally, the starting wavelength, ending wavelength, and constant wavelength interval are specified by "Start: 0.25", "End: 2.5", and "Interval: 0.005", respectively. The material properties are interpolated accordingly to match the wavelength interval provided, allowing the user to leverage computational cost vs. accuracy. Here is an example header with each of these elements used:

```
MC #comment example
Output: test     # output file prefix

Particle 1: TiO2.txt   # load in all material files
Particle 2: BaSO4.txt
Matrix 1: Air.txt
Matrix 2: Acrylic.txt

Solar: am15.txt   # not required
Photons: 30000    # not required if using NN
Start: 0.25       # starting wavelength [microns]
End: 2.5          # ending wavelength [microns]
Interval: 0.005   # wavelength interval [microns]
```

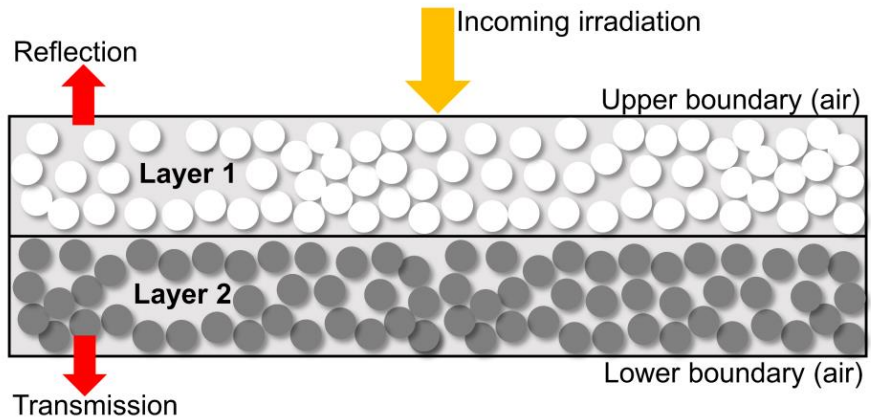

Fig. 2: Schematic based on the example input file of a two layer medium with the incoming irradiation, reflection, transmission, and upper and lower boundaries labelled.

After this, the body is used to provide details about each simulation. Each simulation must be sequentially labelled as "Sim 1", "Sim 2", etc. After labelling the simulation number, the upper and lower refractive index boundary condition can be set by specifying the matrix material it refers to as "Upper: Matrix 1" and "Lower: Matrix 1". If either or both boundary conditions are not set, it will default to air with a refractive index of one ($n = 1$). The refractive index boundary condition is to account for Fresnel reflection at the boundaries. One example where a user may want to set the refractive index boundary is if they simulate paint on an aluminum substrate.

Here the aluminum would be the lower boundary condition. Within each simulation, each layer must be labeled where "Layer 1" is the top layer and any additional layers are added beneath the previous layer. After labeling the layer, one matrix material must be specified such as "Matrix 1" which would reference the material file imported in the header. Each time a matrix is used the thickness must be specified as "T: 100" where all units in the file are based on microns. After the matrix is set, at least one particle must be specified such as "Particle 2" which would reference the material file imported in the header. Each time a particle is used, the diameter (D: 0.4), volume fraction as a percentage (VF: 60), and standard deviation (Std: 0.1) are set. If the standard deviation is not set it will default to 0. Alternatively, for core-shell particles instead of setting the diameter, the core diameter (C: 0.4) and shell thickness (S: 0.1) are set. In section 1 of the supplementary information there are several different simulations highlighting the different features. Below is an example of the body for one simulation with an upper layer of standard nanoparticles and a lower layer of core-shell particles:

```
Sim 1
Upper: Matrix 1
Lower: Matrix 1
Layer 1     # layer 1
Matrix 1
T: 50
Particle 1
D: 0.4      # all units in the input file are in microns
VF: 60
Std: 0.1    # standard deviation (defaults to 0)
Layer 2     # layer 2
Matrix 2
T: 200
Particle 1 # specify material for the core
C: 0.4      # core diameter
Particle 2 # specify material for the shell
S: 0.1      # shell wall thickness
VF: 60      # volume fraction of the core shell particles
```

*Pre-processing*

Generally, it is of interest for the user to simulate photon transport across a wavelength spectrum such as to determine the solar spectral response or the emittance in the sky window. If only one wavelength is of concern, and each of the input material properties are at that wavelength, then no pre-processing is required. However, to simulate across a wavelength range, all the material properties must be at the same wavelengths and intervals which is commonly not the case. To calculate the optical properties, all the material properties must be matched at each wavelength simulated. To handle this, the starting and ending wavelength, and wavelength interval are specified in the header of the input file. FOS then linearly interpolates all material files to match this wavelength range interval. If a material file does not fully cover the specified wavelength range, FOS will warn the user.

*Scattering properties*

It can be a challenging task to calculate the scattering properties for a nanoparticle embedded medium, including the scattering coefficient, absorption coefficient, and asymmetry parameter, from the complex refractive index. Mie theory can be applied for simple geometries, or numerical methods can be used to solve Maxwell's equations for complex geometries. For this program, unless pre-calculated properties are provided, Mie theory is used to calculate the scattering properties for spherical particles with different materials and sizes within the medium as described in Frisvad et al. [14], and implemented by Peoples et al. [1] and Huang et al. [6]. First, the size parameters are calculated by

$$x = \frac{2\pi r \hat{n}_m}{\lambda}, \qquad y = \frac{2\pi r \hat{n}_p}{\lambda} \tag{1}$$

where r is the particle radius, $\hat{n}_m = (n + ik)_m$ is the matrix's complex index of refraction, $\hat{n}_p$ is the nanoparticle's complex index of refraction, and $\lambda$ is the wavelength of light in vacuum [14]. Next, the Mie coefficients are calculated by

$$a_n = \left(\frac{xj_n(x)}{xh_n(x)}\right)\frac{\hat{n}_m A_n(y) - \hat{n}_p A_n(x)}{\hat{n}_m A_n(y) - \hat{n}_p B_n(x)} \tag{2}$$

$$b_n = \left(\frac{xj_n(x)}{xh_n(x)}\right)\frac{\hat{n}_p A_n(y) - \hat{n}_m A_n(x)}{\hat{n}_p A_n(y) - \hat{n}_m B_n(x)} \tag{3}$$

$$A_n(z) = \frac{1}{zj_n(z)}\frac{\partial(zj_n(z))}{\partial z} \tag{4}$$

$$B_n(z) = \frac{1}{zh_n(z)}\frac{\partial(zh_n(z))}{\partial z}. \tag{5}$$

where $j_n(z)$ is the spherical Bessel function of the first kind, and $h_n(z)$ is the spherical Hankel function of the second kind [14]. These functions are implemented as

$$A_n(z) = \frac{1}{2}\sqrt{\frac{\pi}{2z}}\left(\frac{zJ_{n-\frac{1}{2}}(z) + J_{n+\frac{1}{2}}(z) + zJ_{n+\frac{3}{2}}(z)}{\sqrt{\frac{\pi z}{2}}J_{n+\frac{1}{2}}(z)}\right) \tag{6}$$

$$B_n(z) = \frac{1}{2}\sqrt{\frac{\pi}{2z}}\left(\frac{zH_{n-\frac{1}{2}}(z) + H_{n+\frac{1}{2}}(z) + zH_{n+\frac{3}{2}}(z)}{\sqrt{\frac{\pi z}{2}}H_{n+\frac{1}{2}}(z)}\right) \tag{7}$$

where $J_n(z)$ is the ordinary Bessel function of the first kind, and $H_n(z)$ is the ordinary Hankel function of the first kind [48]. The first two Bessel functions at $n = 1, 2$ are calculated using the SciPy library [49], then as $n$ increases ($n \in \mathbb{N}$), the rest of the Bessel function are calculated using recurrence relations given by

$$J_{\nu+1}(z) = \frac{2\nu}{z}J_\nu(z) - J_{\nu-1}(z) \tag{8}$$

$$Y_{\nu+1}(z) = \frac{2\nu}{z} Y_\nu(z) - Y_{\nu-1}(z) \tag{9}$$

where $Y_\nu(z)$ is the Bessel function of the second kind, and the Hankel function of the first kind is calculated as $H_\nu(z) = J_\nu(z) + iY_\nu(z)$ [48]. Using the recurrence relations greatly reduces the computational cost compared to the traditional calculation method for the Bessel functions. Alternatively, for core-shell nanoparticles the Mie coefficients can be calculated following the process described by Bohren et al. [50] and described in section 2 of the supplemental information.

Next, the geometric term ($\gamma$) is calculated to correct for the change in the incident wave due to an absorbing matrix as

$$\gamma = \frac{2(1 + (\alpha - 1)e^\alpha)}{\alpha^2} \tag{10}$$

where $\alpha = \frac{4\pi r}{\lambda k_m}$ [14]. The scattering extinction cross section and scattering cross section of an individual particle are calculated by

$$C_{ext} = \frac{\lambda^2}{2\pi} \sum_{n=1}^{n_{max}} (2n+1) \text{Re}\left(\frac{a_n + b_n}{\hat{n}_m^2}\right) \tag{11}$$

$$C_{sca} = \frac{\lambda^2 \exp(-4\pi r k_{med}/\lambda)}{2\pi\gamma |\hat{n}_m|^2} \sum_{n=1}^{n_{max}} (2n+1)(|a_n|^2 + |b_n|^2) \tag{12}$$

where $n_{max} = \text{ceil}(2 + |y| + 4.3|y|^{1/3})$ is the convergence criteria [14]. The scattering and absorption coefficients of the effective medium is then calculated with effective medium theory by

$$\mu_s = \sum_{i=1}^{\#P} \frac{1.5 q_{sca,i} f_i}{2r} \tag{13}$$

$$\mu_a = \sum_{i=1}^{\#P} \frac{1.5(q_{ext,i} - q_{sca,i})f_i}{2r} \tag{14}$$

where $\#P$ is the number of particle sizes within the medium, $f_i$ is the volume fraction of an individual particle, $q_{sca,i} = C_{sca}/\pi r^2$, and $q_{ext,i} = C_{ext}/\pi r^2$ [18]. The asymmetry parameter of an individual particle is calculated by

$$g_i = \frac{2\sum_{n=1}^{n_{max}} \left( \frac{n(n+2)}{n+1} \text{Re}(a_n a_{n+1}^* + b_n b_{n+1}^*) + \frac{2n+1}{n(n+1)} \text{Re}(a_n b_n^*) \right)}{\sum_{n=1}^{n_{max}}((2n+1)(|a_n|^2 + |b_n|^2))} \tag{15}$$

which is then calculated for the effective medium by

$$g = \frac{1}{\mu_s} \left( \sum_{i=1}^{\#P} \frac{1.5 q_{sca,i} f_i g_i}{2r} \right) \tag{16}$$

[14], [18]. Due to numerical instabilities within the solution process, scattering and absorption coefficients could be less than zero, so these values are corrected by setting them to zero. For nanoparticle volume fractions greater than 0.08 (8%), an approximation is used to account for dependent scattering by

$$C = 1 + 1.5 \left( \sum_{i=1}^{\#P} f_i \right) - 0.75 \left( \sum_{i=1}^{\#P} f_i \right)^2 \tag{17}$$

$$\mu_{a,c} = C\mu_a \tag{18}$$

$$\mu_{s,c} = C\mu_s \tag{19}$$

where $C$ is the correction factor, and $\mu_{a,c}$ and $\mu_{s,c}$ are respectively the corrected absorption and scattering coefficients [51]. This correction does not account for particle clumping or agglomeration. Dependent scattering is a complex phenomenon, so while this correction is a helpful approximation, care should be taken in special cases such as particulate clumping. Finally, the absorption by the matrix is included by

$$\mu_{a,t} = \mu_{a,c} + \frac{4\pi k_m (1 - \sum_{i=1}^{\#P} f_i)}{\lambda} \tag{20}$$

where $\mu_{a,t}$ is the total absorption coefficient for the matrix and the embedded particles [52].

In addition to calculating the effective scattering properties of a medium with multiple different materials or sizes of particles, a standard deviation option is also provided to the user. This is for cases where manufacturing may create a particle size distribution. To account for this, one particle size is converted to 101 different particle sizes following a Gaussian distribution where sizes in the range $D \pm 3\sigma$ are modelled. For custom distributions, one can simply add in the input file as many diameters and volume fractions as required. Distributions are not available for core-shell particles. If the Std setting is not set for a particle in the input file it will default to Std: 0, or no particle size distribution.

*Photon transport*

Photon transport methods are used to calculate the percentage of photons that are reflected, transmitted, or absorbed within a medium. Within this program, there are two available options for calculating photon transport. The first option is a Monte Carlo method that is based on the methods used in MCML by Wang et al. [45]. As for FOS, there is no interest in where photons are absorbed, so the X and Y location of photon absorption is unimportant. To accelerate the program, only the Z-direction and the cosine directional angle off the Z-axis is tracked. Since each photon packet modeled is independent of the others, the photon packets are launched in parallel. This decreases the computational time required by approximately the number of cores used. The refractive index used to account for specular reflectance at the boundaries is that of the matrix, and the scattering coefficient, absorption coefficient, and asymmetry parameter calculated from Mie theory are used to calculate the photon step size, probability of scattering vs. absorption, and new direction angle of the photon. The second available method to estimate photon transport is a machine learning model pre-trained on 50,000 Monte Carlo simulations with 50,000 photons each. This method provides up to 1000-fold speedups over the parallel Monte Carlo simulations for common materials, depending on the thickness and optical properties. Due to the limitations of the neural network model based on the training set, to use this method the input material's properties must be within the ranges shown in Table 1, where only the real part of the refractive index is considered in specular reflection at the boundaries. Additionally, the neural network model can only be used with air or vacuum boundary conditions, whereas the Monte Carlo method can handle any refractive index boundary condition including conductors. Details of the machine learning model can be found in [40]. Both the Monte Carlo method and the machine learning method utilize parallel processing on the CPU. For the Monte Carlo method, the photons run in parallel. This can allow for massive parallelization since the photons are completely independent of each other. For the machine learning method, each wavelength is run in parallel. Typically, between 20 – 300 wavelengths are simulated. On a desktop/workstation computer the number of wavelengths simulated will normally be less than the number of cores allowing for full processor utilization. However, some CPUs have large core counts, some with 128 cores, which may not be fully utilized for the machine learning method if the number of wavelengths simulated is less than the number of cores. For running on computing clusters (supercomputers), FOS is only designed to run in parallel on a single node.

Table 1: Range of optical properties on which the neural network is trained

| Optical Property | Range |
|---|---|
| Matrix Refractive Index (real) | 1-7 |
| Absorption Coefficient (cm$^{-1}$) | 0-1,000,000 |
| Scattering Coefficient (cm$^{-1}$) | 0-200,000 |
| Asymmetry Parameter | 0-1 |
| Medium Thickness (μm) | 5-500 |

*Output*

After modeling the photon transport, if the solar spectrum is included in the input file and the wavelength range covers at least 0.28 – 2.5 μm, the reflectance, absorptance, and transmittance will be integrated to calculate the total properties in the solar spectrum by

$$R_s = \frac{\int_{\lambda_1}^{\lambda_2} R_\lambda G_\lambda d\lambda}{\int_{\lambda_1}^{\lambda_2} G_\lambda d\lambda} \qquad (21)$$

where $R_\lambda$ can be the spectral reflectance, absorptance, or transmittance, and $G_\lambda$ is the spectral solar irradiation [53]. Two output files will be generated per simulation, one with data about the simulation and one with the plotted spectral response. In the file with data, it will include the total solar reflectance, absorptance, and transmittance if the solar file is provided in the input, the spectral response at each wavelength simulated, the scattering properties (refractive index, absorption coefficient, scattering coefficient, asymmetry parameter, and thickness) of each layer at each wavelength, and a copy of the input file information for that simulation at the end for reference. Again, all units in the output file are based on microns.

**Examples**

Four examples are shown to highlight the capabilities and case studies of FOS, as well as to validate it against other open-source programs. First, radiative cooling composites are investigated where a $BaSO_4$-acrylic paint and a $TiO_2$-air particle bed are simulated due to their high solar reflectance and capabilities as radiative cooling materials. Second, water with silicon nanoparticles is simulated as a method to enhance solar absorption in water. Third, a dual layer $Fe_2O_3$-$TiO_2$-acrylic colored paint is shown to highlight the spectral effects of nanoparticle volume fraction and thickness. Fourth, a hollow $SiO_2$ nanoparticle is simulated to show the effect of core-shell nanoparticles in comparison to single material nanoparticles. Each input file for these examples can be found in section 4 of the supplemental information.

*Radiative cooling composites*

Recently, $BaSO_4$ has been found to be an efficient scatterer for radiative cooling paint due to its moderately high bandgap allowing for a good refractive index with no solar absorption, and ample phonon modes enabling high sky window emittance [4], [54]. Three different wavelength intervals are tested including 0.16, 0.04, and 0.01 μm at a 200 μm thickness and 60% volume fraction of 400 nm diameter $BaSO_4$ nanoparticles. The spectral response of three different interval settings can be seen in Fig. 3(A). All three intervals provide reasonable estimates in the 0.5-1.5 μm wavelength spectrum as there is little variation in this region, however the different interval settings deviate in regions of high variation such as the 1.6-1.9 μm wavelength spectrum. Additionally, the 0.16 μm interval provides a similar solar reflectance as the 0.04 μm interval (0.842 vs. 0.847) even though Fig. 3(A) shows there are large reflectance differences between the interval settings. This is due to the errors partially cancelling with the larger wavelength interval, where in some areas reflectance is overpredicted (such as from 2.25-2.5 µm) and in other areas reflectance is underpredicted (such as from 0.25-0.5 µm). While solar reflectance is a good convergence criterion, we also recommend visually confirming in the plotted spectral reflectance that the wavelength interval is fine enough to capture the features in the spectral response. In Fig. 3(B) the reflectance (R), absorptance (A), and transmittance (T) predictions from the pre-trained neural network are compared to Monte Carlo simulations where the maximum absolute error at a single wavelength is 0.0105 and the average error across all wavelengths is 0.0018. The acceptable error is completely dependent on the user and application. For this example, the neural network approximation takes roughly 3 seconds while the Monte Carlo simulation at 50,000 photons (the same number of photons the neural network is trained on) takes 28 seconds on a desktop computer providing a speedup of 9.3x. However, the majority of the neural network computational time is spent compiling the feed-forward function. When running 20 simulations the neural network still only takes roughly 3 seconds while the Monte Carlo simulation takes 775 seconds providing a speedup of 258x. Since the compilation is a one-time event, the more simulations run the higher the speedup will be. While the neural network is trained on a wide range of optical properties, the error should always be evaluated next to Monte Carlo simulations before use for high throughput screening or optimization. With the pretrained neural network and the wavelength interval control abilities available, high throughput screening is now possible to search for potential materials, as well as ideal nanoparticle sizes and volume fractions. To validate the Mie theory and Monte Carlo implementation used in FOS, Figs 3(C-D) show a comparison against other solvers with 500 nm $TiO_2$ nanoparticles in air at a 5% volume fraction and 100 µm thickness. Figure 3(C) compares the scattering and extinction efficiencies calculated through Mie theory against the miepython open-source code by Scott Prahl [42]. The efficiencies calculated by FOS match precisely to those through miepython. Figure 3(D) compared FOS's Monte Carlo implementation to the well-known MCML by Wang et al [45]. Here we see FOS's spectral response prediction has an average error of 0.0004 compared to MCML with 300,000 photons.

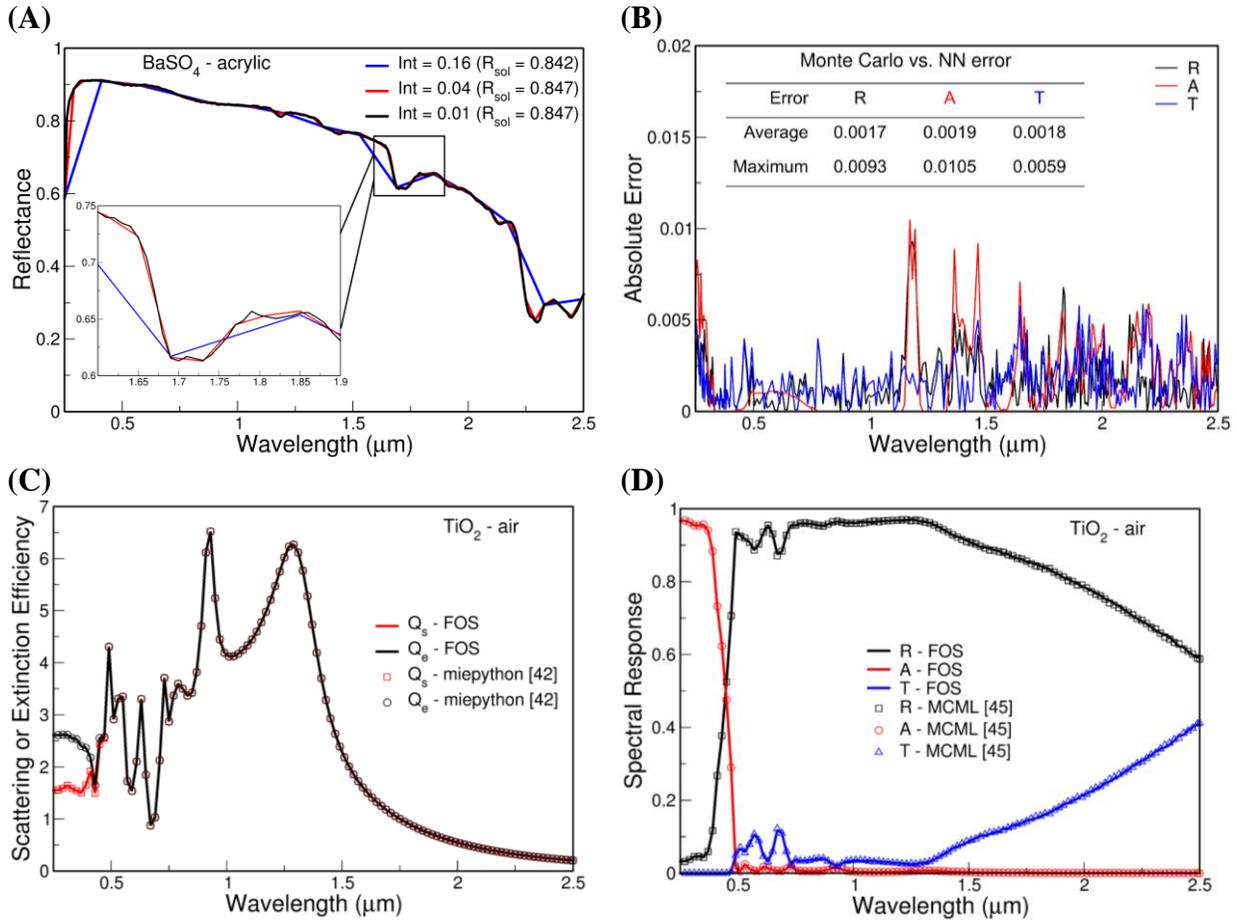

Fig. 3: **(A)** Reflectance as a function of wavelength for four different mesh settings of 400 nm $BaSO_4$ at a 60% volume fraction in a 200 μm acrylic layer. **(B)** absolute error as a function of wavelength comparing the Monte Carlo simulation to the neural network predicted reflectance (R), absorptance (A) and transmittance (T). **(C)** Scattering and extinction efficiency calculated by FOS compared to miepython by Scott Prahl of a 5% volume fraction of 500 nm $TiO_2$ particles in air. **(D)** Spectral response prediction by FOS compared to MCML by Wang et al. of 500 nm $TiO_2$ particles in air at a 5% volume fraction and 100 μm thickness.

*Si-water*

Another use case FOS could be applied to is solar heating of water by adding silicon nanoparticles. Work by Ishii et al. shows specially sized silicon nanoparticles at Mie resonances can enhance solar heating to increase vaporization rates [7]. In Fig. 4(A) three different particle sizes are shown and in Fig. 4(B) three different particle size distributions are shown. As the diameter changes, the Mie resonances shift the absorption peaks. A size distribution cancels out these peaks and approaches a smooth line, which may or may not be beneficial dependent on the application such as for coloration. This highlights the importance of modeling nanoparticle size distributions. In cases such as this where there is a large difference in refractive index between the nanoparticle and the matrix, or if the refractive index varies significantly, then a particle size distribution will play a major role. In cases where the refractive index difference is small and the refractive index does not change significantly, then a distribution may not have a significant impact.

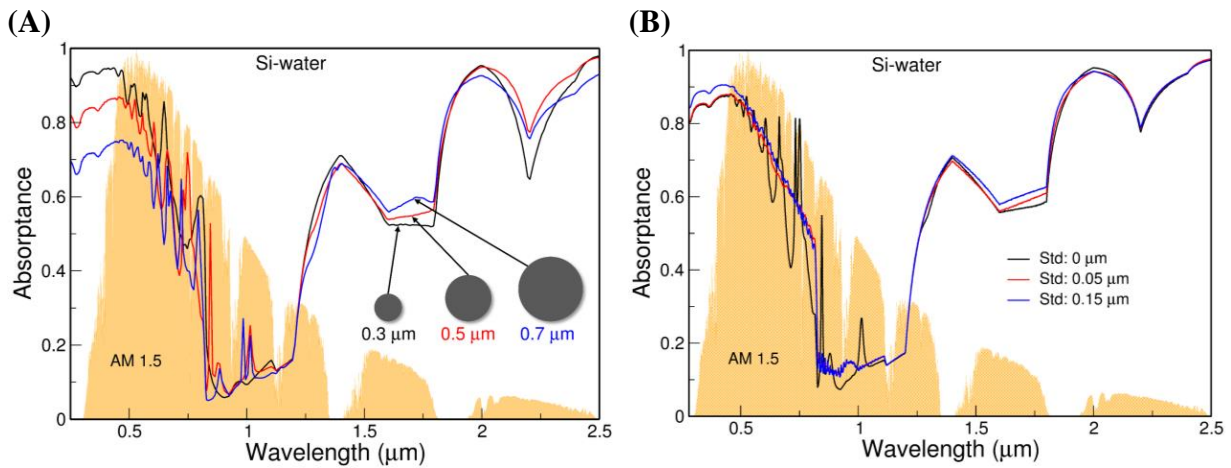

Fig. 4: Absorptance as a function of wavelength for a 0.1% volume fraction of silicon nanoparticles in 500 μm of water for **(A)** three different particle sizes and **(B)** three particle size distributions.

*Fe$_2$O$_3$-TiO$_2$-acrylic*

For this example, the multi-layer and multi-particle capabilities are highlighted. Here, a top layer of Fe$_2$O$_3$-acrylic paint is placed on top of a layer of TiO$_2$-acrylic paint. It should be noted that Fe$_2$O$_3$ is an anisotropic material with a different refractive index based on the crystal orientation. One can average the refractive index for an approximation, particularly if the values are similar. This type of bilayer design, shown in Fig. 5(A), is a commonly used method for creating colored radiative cooling paints as the short solar wavelengths will interact with the top color pigment layer, and the long solar wavelengths are more likely to pass through and interact with the bottom layer. As seen in Figs. 5(B)-5(C), the volume fraction and thickness of the top layer can significantly impact the paint's color due to the high extinction coefficient of Fe$_2$O$_3$. The paint color is calculated from the spectral reflectance. FOS can model multi-layer mediums as well as multiple nanoparticles within each layer. This will allow for accelerated development of colored radiative cooling paints by allowing researchers to search for materials, methods, and particle sizes to provide color while maximizing solar reflectance. This can also allow researchers to combine particle types to maximize both sky window emission and solar reflectance and be able to determine the right balance between these two goals.

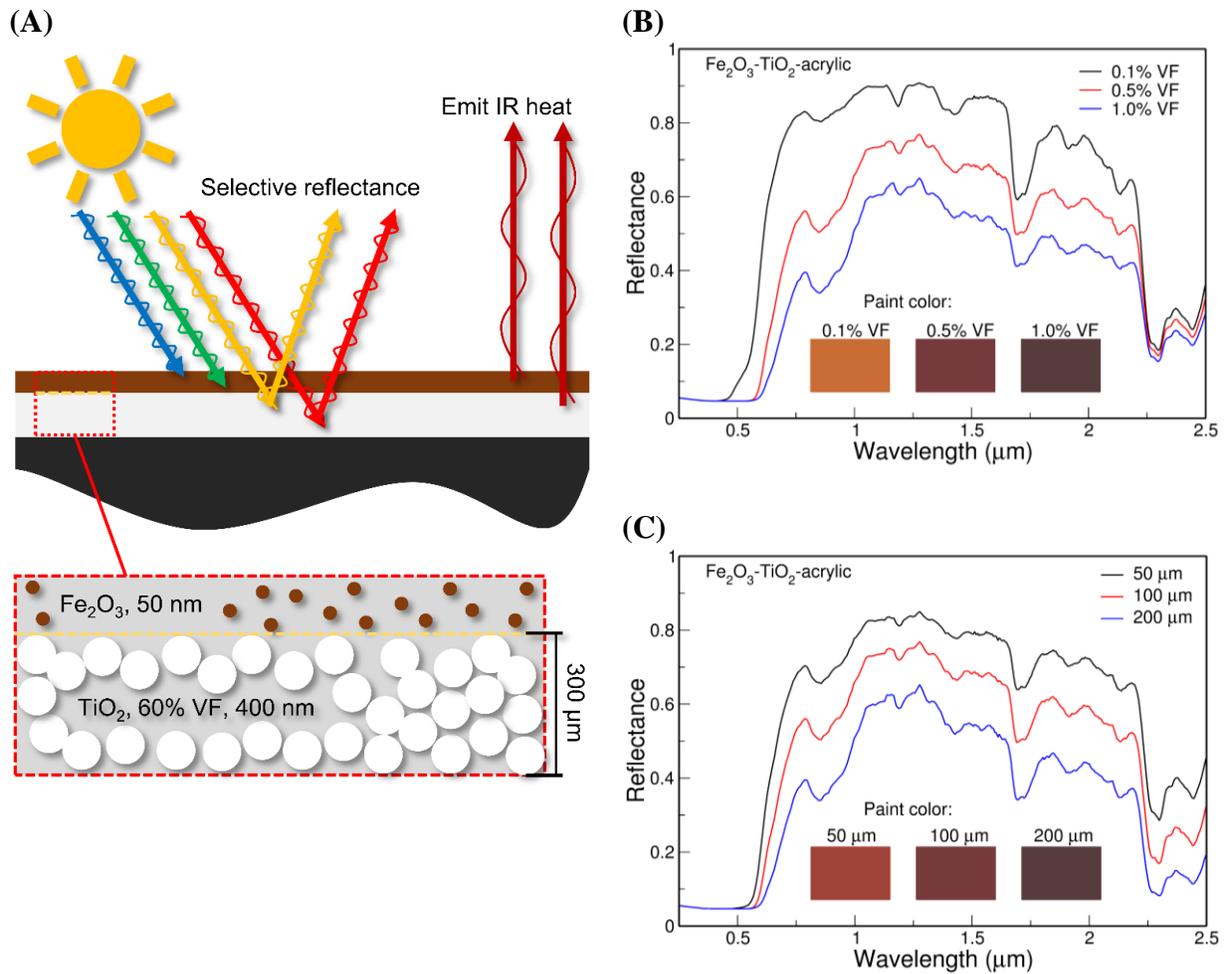

Fig. 5: **(A)** Sketch of a bilayer radiative cooling paint; reflectance as a function of wavelength for **(B)** a bilayer paint with a 100 µm top layer of 50 nm $Fe_2O_3$ particles in acrylic at three different volume fractions and a 300 µm bottom layer of 400 nm diameter $TiO_2$ particles in acrylic at a 60% volume fraction and **(C)** a bilayer paint with a top layer of 50 nm diameter $Fe_2O_3$ particles in acrylic at a 0.5% volume fraction at three different thicknesses and a 300 µm bottom layer of 400 nm diameter $TiO_2$ particles in acrylic at a 60% volume fraction.

*Hollow $SiO_2$*

For the last example, we will demonstrate core-shell nanoparticle modeling. Core-shell nanoparticles are capable of providing unique optical properties for radiative cooling as well as creating sharp absorption peaks [55], [56]. Here, a 50 µm layer with a 60% volume fraction of hollow SiO$_2$ nanoparticles in air is fixed at an inner core diameter of 0.5 µm while the shell wall thickness is tested at three different values as shown in Fig. 6(A). Care should be taken here as for thin shells or particles less than 20 nm, size dependent properties may play a large role and the refractive index should be adjusted accordingly before running FOS [57]. Additionally for comparison in Fig. 6(B), a 50 µm layer with a 60% volume fraction of solid SiO$_2$ nanoparticles in air are modeled at the same total outer diameter as in Fig. 6(A). Here, we see that in the shorter wavelengths, the reflectance is similar, but in the longer wavelengths the hollow core-shell particle reflectance drops off. This could be beneficial for spectral selectivity, or for absorbing nanoparticles, the core-shell dimensions can be tuned to create an absorption peak. Alternatively, dependent on the core-shell dimensions, reflectance could also be boosted greater than that of a solid particle.

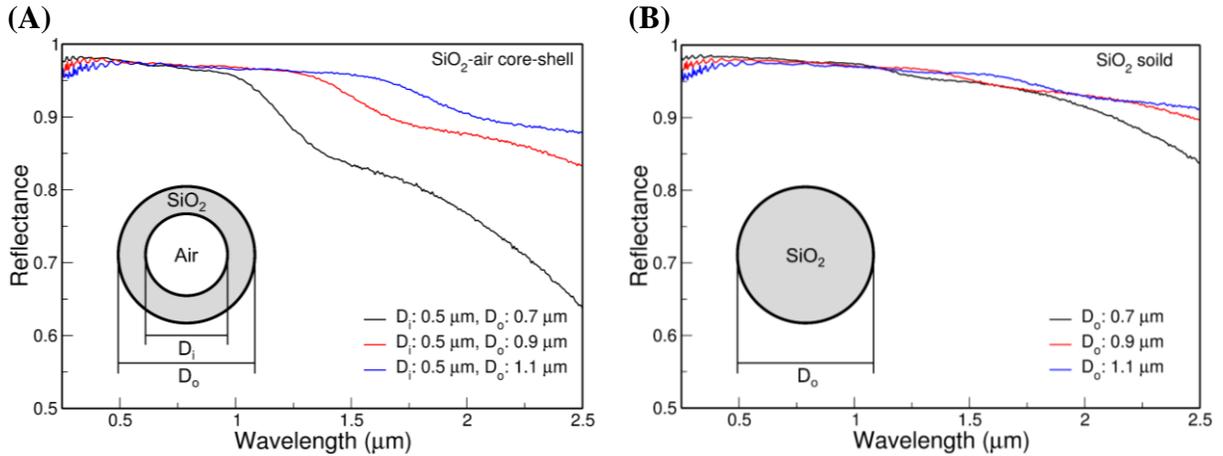

Fig. 6: Reflectance as a function of wavelength for **(A)** a 50 µm layer of 0.5 µm inner diameter hollow SiO$_2$ nanoparticles at a 60% volume fraction with outer diameter ranging from 0.7 – 1.1 µm, and **(B)** a 50 µm layer of solid SiO$_2$ nanoparticles at a 60% volume fraction with outer diameter ranging from 0.7 – 1.1 µm.

**Conclusion**

        Modeling tools combining Mie theory with photon transport simulations are critical in simulating the spectral response of nanoparticulate media for applications like radiative cooling paints, dust coating on radiators, and solar heating of water. FOS provides two main contributions to the field. First, it combines all the necessary methods together into one easy-to-use program which will remove the barrier of entry into simulations. This will provide value to research groups that focus on experimental work but wish to run photon transport calculations. Second, it gives researchers access to computationally efficient methods, such as integrated Mie theory with parallel Monte Carlo simulations, interpolation and mesh reduction methods, and a pre-trained neural network to directly replace Monte Carlo simulations for accelerated photon transport predictions. With these tools, FOS will allow for efficient calculation of nanoparticle radiative properties, accelerated high throughput design, and expedited optimization of particle sizes, volume fractions, and particle material and size combinations. This could accelerate the development of radiative cooling paints, including the use of colored or fluorescent pigments, help address the challenges of lunar dust modeling for spacecraft radiators, and increase efficiency across many different fields using these methods.

## Declaration of Competing Interest

The authors declare no competing interest with the work presented.

## Acknowledgments

We acknowledge partial support from the US National Science Foundation (Award #2102645). J.P. acknowledges support from the NASA Space Technology Graduate Research Opportunity Program (Grant #80NSSC20K1187).

# References

[1] J. Peoples, X. Li, Y. Lv, J. Qiu, Z. Huang, and X. Ruan, "A strategy of hierarchical particle sizes in nanoparticle composite for enhancing solar reflection," *Int. J. Heat Mass Transfer*, vol. 131, pp. 487–494, 2019, doi: 10.1016/j.ijheatmasstransfer.2018.11.059.

[2] A. Felicelli *et al.*, "Thin layer lightweight and ultrawhite hexagonal boron nitride nanoporous paints for daytime radiative cooling," *Cell Reports Physical Science*, vol. 3, p. 101058, 2022.

[3] Md. M. Hossain and M. Gu, "Radiative cooling: principles, progress, and potential," *Adv. Sci.*, vol. 3, p. 1500360, 2016.

[4] X. Li, J. Peoples, P. Yao, and X. Ruan, "Ultrawhite BaSO4 Paints and Films for Remarkable Daytime Subambient Radiative Cooling," *ACS Appl. Mater. Interfaces*, vol. 13, no. 18, pp. 21733–21739, Apr. 2021, doi: 10.1021/acsami.1c02368.

[5] R. A. Yalçın, E. Blandre, K. Joulain, and J. Drévillon, "Colored Radiative Cooling Coatings with Nanoparticles," *ACS Photonics*, vol. 7, no. 5, pp. 1312–1322, Apr. 2020, doi: 10.1021/acsphotonics.0c00513.

[6] Z. Huang and X. Ruan, "Nanoparticle embedded double-layer coating for daytime radiative cooling," *Int. J. Heat Mass Transfer*, vol. 104, pp. 890–896, 2017, doi: 10.1016/j.ijheatmasstransfer.2016.08.009.

[7] S. Ishii, R. P. Sugavaneshwar, K. Chen, T. D. Dao, and T. Nagao, "Solar water heating and vaporization with silicon nanoparticles at mie resonances," *Opt. Mater. Express*, vol. 6, pp. 640–648, 2016.

[8] Y.-C. Lu and C.-H. Hsueh, "Subwavelength VO2 Nanoparticle Films for Smart Window Applications," *ACS Appl. Nano Mater.*, vol. 5, pp. 2923–2934, 2022.

[9] Y. Ke *et al.*, "Cephalopod-inspired versatile design based on plasmonic VO2 nanoparticle for energy-efficient mechano-thermochromic windows," *Nano Energy*, vol. 73, p. 104785, 2020.

[10] N. M. Mohammad *et al.*, "Highly tunable cellulosic hydrogels with dynamic solar modulation for energy-efficient windows," *Small*, p. 2303706, 2024.

[11] X. Huang and M. A. El-Sayed, "Gold nanoparticles: Optical properties and implementations in cancer diagnosis and photothermal therapy," *Journal of Advanced Research*, vol. 1, no. 1, pp. 13–28, 2010.

[12] N. G. Khlebtsov and L. A. Dykman, "Optical properties and biomedical applications of plasmonic nanoparticles," *Journal of Quantitative Spectroscopy and Radiative Transfer*, vol. 111, no. 1, pp. 1–35, 2010.

[13] D. M. Sullivan, *Electromagnetic simulation using the FDTD method*. John Wiley & Sons, 2013.

[14] J. R. Frisvad, N. J. Christensen, and H. W. Jensen, "Computing the scattering properties of participating media using Lorenz-Mie theory," *ACM Trans. Graph.*, vol. 26, no. 3, pp. 60-es, Jul. 2007, doi: 10.1145/1276377.1276452.

[15] M. Kahnert, "Numerical methods in electromagnetic scattering theory," *Journal of Quantitative Spectroscopy and Radiative Transfer*, vol. 79–80, pp. 775–824, 2003, doi: 10.1016/S0022-4073(02)00321-7.

[16] Q. Fu and W. Sun, "Mie theory for light scattering by a spherical particle in an absorbing medium," *Appl. Opt.*, vol. 40, pp. 1354–1361, 2001.

[17] W. C. Mundy, J. A. Roux, and A. M. Smith, "Mie scattering by spheres in an absorbing medium," *J. Opt. Soc. Am.*, vol. 64, pp. 1593–1597, 1974.

[18] M. F. Modest and S. Mazumder, *Radiative Heat Transfer*. Elsevier Science, 2013.
[19] P. Kubelka and F. Munk, "An article on optics of paint layers," *Z. Tech. Phys*, vol. 12, pp. 259–274, 1931.
[20] "Revised Kubelka–Munk theory. I. Theory and application," *J. Opt. Soc. Am. A*, vol. 21, no. 10, pp. 1933–1941, 2004.
[21] P. Kubelka, "New contributions to the optics of intensely scattering materials. Part I," *J. Opt. Soc. Am.*, vol. 38, pp. 448–457, 1948.
[22] H. C. Hottel and E. S. Cohen, "Radiant heat exchange in a gas-filled enclosure: Allowance for nonuniformity of gas temperature," *AIChE Journal*, vol. 4, no. 1, pp. 3–14, 1958.
[23] W. W. Yuen and E. E. Takara, "Development of a generalized zonal method for analysis of radiative transfer in absorbing and anisotropically scattering media," *Numerical Heat Transfer, Part B Fundamentals*, vol. 25, no. 1, pp. 75–96, 1994.
[24] S. Chandrasekhar, *Radiative Transfer*. Courier Corporation, 2013.
[25] P. J. Coelho, "Advances in the discrete ordinates and finite volume methods for the solution of radiative heat transfer problems in participating media," *Journal of Quantitative Spectroscopy and Radiative Transfer*, vol. 145, pp. 121–146, 2014.
[26] S. A. Prahl, M. van Gemert, and A. Welch, "Determining the optical properties of turbid media by using the adding-doubling method," *Appl. Opt.*, vol. 32, no. 4, pp. 559–568, 1993.
[27] A. Welch and M. van Gemert, "The Adding-Doubling Method. In: Optical-Thermal Response of Laser-Irradiated Tissue. Lasers, Photonics, and Electro-Optics," Springer, 1995.
[28] J. R. Howell and M. Perlmutter, "Monte Carlo Solution of Thermal Transfer Through Radiant Media Between Gray Walls," *ASME. J. Heat Transfer.*, vol. 86, no. 1, pp. 116–122, Feb. 1964, doi: 10.1115/1.3687044.
[29] J. R. Howell, "The Monte Carlo Method in Radiative Heat Transfer," *ASME J. Heat Transfer*, vol. 120, no. 3, pp. 547–560, 1998.
[30] J. Farmer and J. R. Howell, "Comparison of Monte Carlo Strategies for Radiative Transfer in Participating Media," *Advances in Heat Transfer*, vol. 31, pp. 333–429, 1998.
[31] A. Haghighat and J. Wagner, "Monte Carlo variance reduction with deterministic importance functions," *Progress in Nuclear Energy*, vol. 42, no. 1, pp. 25–53, 2003.
[32] I. Kawrakow and M. Fippel, "Investigation of variance reduction techniques for Monte Carlo photon dose calculation using XVMC," *Phys. Med. Biol.*, vol. 45, p. 2163, 2000.
[33] J. R. Howell and K. J. Daun, "The Past and Future of the Monte Carlo Method in Thermal Radiation Transfer," *ASME J. Heat Transfer*, vol. 143, no. 10, p. 100801, 2021.
[34] Z. Guo *et al.*, "Fast and accurate machine learning prediction of phonon scattering rates and lattice thermal conductivity," *npj Comput. Mater*, vol. 9, p. 95, 2023.
[35] D. Kochkov, J. A. Smith, A. Alieva, Q. Wang, M. P. Brenner, and S. Hoyer, "Machine learning–accelerated computational fluid dynamics," *PNAS*, vol. 118, no. 21, p. e2101784118, 2021.
[36] H. Wei, H. Bao, and X. Ruan, "Perspective: Predicting and optimizing thermal transport properties with machine learning methods," *Energy and AI*, vol. 8, p. 1000153, 2022.
[37] H. H. Kang, M. Kaya, and S. Hajimirza, "A data driven artificial neural network model for predicting radiative properties of metallic packed beds," *Journal of Quantitative Spectroscopy and Radiative Transfer*, vol. 226, pp. 66–72, 2019.

[38] P. G. Stegmann, B. Johnson, I. Moradi, B. Karpowicz, and W. McCarty, "A deep learning approach to fast radiative transfer," *Journal of Quantitative Spectroscopy and Radiative Transfer*, vol. 280, p. 108088, 2022.

[39] A. Royer, O. Farges, P. Boulet, and D. Burot, "A new method for modeling radiative heat transfer based on Bayesian artificial neural networks and Monte Carlo method in participating media," *Int. J. Heat Mass Transfer*, vol. 201, no. Part 1, p. 123610, 2023.

[40] D. Carne, J. Peoples, D. Feng, and X. Ruan, "Accelerated Prediction of Photon Transport in Nanoparticle Media Using Machine Learning Trained With Monte Carlo Simulations," *ASME J. Heat Mass Transfer*, vol. 145, no. 5, p. 052502, 2023.

[41] X. Li, J. Peoples, Z. Huang, Z. Zhao, J. Qiu, and X. Ruan, "Full Daytime Sub-ambient Radiative Cooling in Commercial-like Paints with High Figure of Merit," *Cell Reports Physical Science*, vol. 1, no. 10, p. 100221, 2020, doi: 10.1016/j.xcrp.2020.100221.

[42] S. Prahl, "miepython: Pure python calculation of Mie scattering." May 07, 2024.

[43] B. Sumlin, W. Heinson, and R. Chakrabarty, "Retrieving the aerosol complex refractive index using PyMieScatt: A Mie computational package with visualization capabilities," *Journal of Quantitative Spectroscopy and Radiative Transfer*, vol. 205, pp. 127–134, 2018.

[44] C. Mätzler, "MATLAB functions for Mie scattering and absorption." 2002.

[45] L. Wang, S. L. Jacques, and L. Zheng, "MCML—Monte Carlo modeling of light transport in multi-layered tissues," *Comput. Methods Programs Biomed.*, vol. 47, no. 2, pp. 131–146, 1995, doi: 10.1016/0169-2607(95)01640-F.

[46] P. Romano, N. Horelik, B. Herman, A. Nelson, B. Forget, and K. Smith, "OpenMC: A state-of-the-art Monte Carlo code for research and development," *Annals of Nuclear Energy*, vol. 82, pp. 90–97, 2015.

[47] D. Marti, R. Aasbjerg, P. Andersen, and A. Hansen, "MCmatlab: an open-source, user-friendly, MATLAB-integrated three-dimensional Monte Carlo light transport solver with heat diffusion and tissue damage," *Journal of Biomedical Optics*, vol. 23, no. 12, p. 121622, 2018.

[48] M. Abramowitz and I. Stegun, *Handbook of mathematical functions with formulas, graphs, and mathematical tables*. United States Department of Commerce, National Bureau of Standards, 1964.

[49] E. Jones, T. Oliphant, P. Peterson, and others, "Scipy: Open source scientific tools for Python." 2001. [Online]. Available: http://www.scipy.org/

[50] C. F. Bohren and D. R. Huffman, "A Potpourri of Particles," in *Absoprtion and Scattering of Light by Small Particles*, Wiley, 1998, pp. 181–223.

[51] M. Kaviany, *Principles of Heat Transfer in Porous Media*, 2nd ed. New York: Springer, 1995.

[52] D. Carne, J. Peoples, F. Arentz, and X. Ruan, "True benefits of multiple nanoparticle sizes in radiative cooling paints identified with machine learning," *International Journal of Heat and Mass Transfer*, vol. 222, p. 125209, 2024.

[53] T. Bergman, A. Lavine, F. Incropera, and D. DeWitt, *Fundamentals of heat and mass transfer*, 8th ed. Wiley, 2018.

[54] Z. Tong, J. Peoples, X. Li, X. Yang, H. Bao, and X. Ruan, "Electronic and Phononic Origins of BaSO4 as an Ultra-Efficient Radiative Cooling Paint Pigment," vol. 24, p. 100658, 2022.

[55] D. Hu, S. Sun, P. Du, X. Lu, H. Zhang, and Z. Zhang, "Hollow Core-Shell Particle-Containing Coating for Passive Daytime Radiative Cooling," *Composites: Part A*, vol. 185, p. 106949, 2022.
[56] R. A. Yalcin, E. Blandre, K. Joulain, and J. Drevillon, "Colored Radiative Cooling Coarings with Nanoparticles," *ACS Photonics*, vol. 7, no. 5, 2020.
[57] L. Herrera, D. Arboleda, D. Schinca, and L. Scaffardi, "Determination of plasma frequency, damping constant, and size distribution from the complex dielectric function of noble metal nanoparticles," *Journal of Applied Physics*, vol. 116, p. 233105, 2014.

# Supplemental Information for:

# FOS: A fully integrated open-source program for Fast Optical Spectrum calculations for nanoparticle media

## 1: Example input file

```
MC
# MC for Monte-Carlo or NN for neural network, must be in the first line

# hashtags comment a line out
# output defines the output files prefix
Output: Test  # comments can also be next to input

# import particle and matrix material files which will be used in the simulations below
# materials files need three columns, wavelength, refractive index, and extinction coefficient
Particle 1: BaSO4.txt
Particle 2: Si.txt
Matrix 1: acr.txt
Matrix 2: air.txt

# if solar is included, the output file will include the integrated solar response
Solar: am15.txt

# number of photons per wavelength for Monte Carlo simulations
Photons: 30000

# wavelength range and interval in microns
Start: 0.25
End: 2.5
Interval: 0.005

### End header ###
### Start body ###

# in the body you can add as many sims as you would like

# label each simulation starting with sim: 1
Sim 1
# upper and lower represent the upper and lower refractive index boundaries
# if either or both are left out, it will default to air (n=1), or can be specified as one of the matrix materials
Upper: Matrix 2
Lower: Matrix 2
# declare each layer number before inputting the details of the layer
Layer 1
# after declaring a layer, one medium and thickness must be set
Matrix 1
T: 500 # all input units in microns
# after defining the medium, add as many particle materials as you need
Particle 1
# for each particle, the diameter (D), volume fraction (VF), and distribution (Dist) must be set
D: 0.5 # microns
```

```
VF: 58 #percentage
Std: 0.1      # Standard deviation of a particle size. If left out, will default to 0
Particle 2
D: 0.08
VF: 0.5
Std: 0

## this is an example of a 2 layer simulation
Sim 2
# boundary conditions are left out here, which default to air
Layer 1
Matrix 1
T: 100
Particle 1
D: 0.4
VF: 30
Layer 2
Matrix 1
T: 400
Particle 2
D: 0.5
VF: 60
Std: 0.1

## this is an example of a core shell simulation
Sim 3
Layer 1
Matrix 1
T: 100
Particle 1
C: 0.4 # core diameter
Particle 2
S: 0.01      # shell wall thickness
VF: 30
Std: 0 # Std must be 0 for core shell

## this is an example of multiple particle sizes of a single material
Sim 4
Layer 1
Matrix 1
T: 100
Particle 1
# number of diameters, volume fractions, and standard deviations must match
D: 0.1, 0.2, 0.3, 0.4, 0.5
VF: 1, 5, 10, 5, 1
Std: 0.01, 0.02, 0.03, 0.04, 0.05
Particle 2
D: 0.8, 0.9
VF: 4, 8

## this is an example of an upper layer with core shell particles with a lower layer of standard particles
Sim 5
Layer 1
Matrix 1
T: 100
Particle 1
C: 0.4
Particle 2
S: 0.01
VF: 30
```

```
Layer 2
Matrix 2
T: 1000
Particle 1
D: 0.4
VF: 20
Particle 2
D: 0.5
VF: 30
```

## 2: Example material files

If entering the refractive index, the material file will consist of three columns including the wavelength, the refractive index and extinction coefficient at each wavelength.
Example:
0.250   1.55   6.50e-05
0.253   1.55   2.50e-05
0.257   1.54   1.20e-05
0.261   1.54   7.00e-06
0.265   1.54   5.50e-06
0.269   1.54   4.65e-06
0.273   1.54   4.03e-06
0.276   1.53   3.36e-06
0.280   1.53   2.77e-06
0.284   1.53   2.24e-06
0.288   1.53   1.77e-06
0.292   1.53   1.44e-06
0.296   1.52   1.19e-06

If entering the pre-calculated scattering properties, the material file will consist of four columns including the wavelength, the absorption coefficient, scattering coefficient, and asymmetry parameter at each wavelength where all units are based in µm.
Example:
0.250   0.152   20.5   0.516
0.253   0.146   30.8   0.510
0.257   0.139   38.4   0.503
0.261   0.137   45.8   0.499
0.265   0.136   52.1   0.498
0.269   0.135   59.3   0.498
0.273   0.134   65.7   0.497
0.276   0.134   69.6   0.496
0.280   0.134   71.0   0.494
0.284   0.133   72.8   0.491
0.288   0.133   72.5   0.485
0.292   0.133   71.4   0.472
0.296   0.133   70.9   0.467

## 3: Core-Shell Mie theory equations

The Mie coefficients, $a_n$ and $b_n$, are calculated for core-shell particles by

$$a_n = \frac{\psi_n(y)[\psi'_n(m_2 y) - A_n \chi'_n(m_2 y)] - m_2 \psi'_n(y)[\psi_n(m_2 y) - A_n \chi_n(m_2 y)]}{\xi_n(y)[\psi'_n(m_2 y) - A_n \chi'_n(m_2 y)] - m_2 \xi'_n(y)[\psi_n(m_2 y) - A_n \chi_n(m_2 y)]} \quad \text{(S1)}$$

$$b_n = \frac{m_2 \psi_n(y)[\psi'_n(m_2 y) - B_n \chi'_n(m_2 y)] - \psi'_n(y)[\psi_n(m_2 y) - B_n \chi_n(m_2 y)]}{m_2 \xi_n(y)[\psi'_n(m_2 y) - B_n \chi'_n(m_2 y)] - \xi'_n(y)[\psi_n(m_2 y) - B_n \chi_n(m_2 y)]} \quad \text{(S2)}$$

where

$$A_n = \frac{m_2 \psi_n(m_2 x)\psi'_n(m_1 x) - m_1 \psi'_n(m_2 x)\psi_n(m_1 x)}{m_2 \chi_n(m_2 x)\psi'_n(m_1 x) - m_1 \chi'_n(m_2 x)\psi_n(m_1 x)} \quad \text{(S3)}$$

$$B_n = \frac{m_2 \psi_n(m_1 x)\psi'_n(m_2 x) - m_1 \psi_n(m_2 x)\psi'_n(m_1 x)}{m_2 \chi'_n(m_2 x)\psi_n(m_1 x) - m_1 \psi'_n(m_1 x)\chi_n(m_2 x)} \quad \text{(S4)}$$

$$m_1 = \frac{(n + ik)_{core}}{(n + ik)_m} \quad \text{(S5)}$$

$$m_2 = \frac{(n + ik)_{shell}}{(n + ik)_m} \quad \text{(S6)}$$

$$\psi_n(x) = \sqrt{\frac{\pi x}{2}} J_{n+\frac{1}{2}}(x) \quad \text{(S7)}$$

$$\chi_n(x) = \sqrt{\frac{\pi x}{2}} Y_{n+\frac{1}{2}}(x) \quad \text{(S8)}$$

$$\xi_n(x) = \sqrt{\frac{\pi x}{2}} H_{n+\frac{1}{2}}(x). \quad \text{(S9)}$$

## 4: Input files from examples

For Fig. 3(A) 4 input different files are needed to change the mesh setting, one is shown here.

**Input file from Fig. 3(A):**

```
MC
Output: fig3a
Particle 1: baso4.txt
Matrix 1: acr.txt
Solar: AM15.txt
Photons: 100000
Start: 0.25
End: 2.5
interval: 0.005

Sim 1
Layer 1
Matrix 1
```

```
T: 200
Particle 1
D: 0.4
VF: 60
```

**Input file from Fig. 3(B):**
```
NN
Output: fig3b
Particle 1: baso4.txt
Matrix 1: acr.txt
Solar: AM15.txt
Photons: 100000
Start: 0.25
End: 2.5
interval: 0.005

Sim 1
Layer 1
Matrix 1
T: 200
Particle 1
D: 0.4
VF: 60
```

**Input file from Fig. 4(A):**
```
MC
Output: fig4a
Particle 1: si.txt
Matrix 1: water.txt
Solar: AM15.txt
Photons: 100000
Start: 0.25
End: 2.5
interval: 0.005

Sim 1
Layer 1
Matrix 1
T: 500
Particle 1
D: 0.3
VF: 0.1

Sim 2
Layer 1
Matrix 1
T: 500
Particle 1
D: 0.5
VF: 0.1

Sim 3
Layer 1
Matrix 1
T: 500
Particle 1
D: 0.7
VF: 0.1
```

**Input file from Fig. 4(B):**

```
MC
Output: fig4b
Particle 1: si.txt
Matrix 1: water.txt
Solar: AM15.txt
Photons: 100000
Start: 0.25
End: 2.5
interval: 0.005

Sim 1
Layer 1
Matrix 1
T: 500
Particle 1
D: 0.5
VF: 0.1

Sim 2
Layer 1
Matrix 1
T: 500
Particle 1
D: 0.5
VF: 0.1
Std: 0.05

Sim 3
Layer 1
Matrix 1
T: 500
Particle 1
D: 0.5
VF: 0.1
Std: 0.15
```
**Input file from Fig. 5(A):**
```
MC
Output: fig5a
Particle 1: fe2o3.txt
Particle 2: tio2.txt
Matrix 1: acr.txt
Solar: AM15.txt
Photons: 100000
Start: 0.25
End: 2.5
interval: 0.005

Sim 1
Layer 1
Matrix 1
T: 100
Particle 1
D: 0.05
VF: 0.1
Layer 2
Matrix 1
T: 300
Particle 2
D: 0.4
VF: 60
```

```
Sim 2
Layer 1
Matrix 1
T: 100
Particle 1
D: 0.05
VF: 0.5
Layer 2
Matrix 1
T: 300
Particle 2
D: 0.4
VF: 60

Sim 3
Layer 1
Matrix 1
T: 100
Particle 1
D: 0.05
VF: 1.0
Layer 2
Matrix 1
T: 300
Particle 2
D: 0.4
VF: 60
```
**Input file from Fig. 5(B):**
```
MC
Output: fig5b
Particle 1: fe2o3.txt
Particle 2: tio2.txt
Matrix 1: acr.txt
Solar: AM15.txt
Photons: 100000
Start: 0.25
End: 2.5
interval: 0.005

Sim 1
Layer 1
Matrix 1
T: 50
Particle 1
D: 0.05
VF: 0.5
Layer 2
Matrix 1
T: 300
Particle 1
D: 0.4
VF: 60

Sim 2
Layer 1
Matrix 1
T: 100
Particle 1
D: 0.05
VF: 0.5
```

```
Layer 2
Matrix 1
T: 300
Particle 1
D: 0.4
VF: 60

Sim 3
Layer 1
Matrix 1
T: 200
Particle 1
D: 0.05
VF: 0.5
Layer 2
Matrix 1
T: 300
Particle 1
D: 0.4
VF: 60
```
**Input file from Fig. 6(A):**
```
MC
Output: fig6a
Particle 1: sio2.txt
Particle 2: air.txt
Matrix 1: air.txt
Solar: AM15.txt
Photons: 100000
Start: 0.25
End: 2.5
interval: 0.005

Sim 1
Layer 1
Matrix 1
T: 50
Particle 2
C: 0.5
Particle 1
S: 0.1
VF: 60

Sim 2
Layer 1
Matrix 1
T: 50
Particle 2
C: 0.5
Particle 1
S: 0.2
VF: 60

Sim 3
Layer 1
Matrix 1
T: 50
Particle 2
C: 0.5
Particle 1
S: 0.3
VF: 60
```

**Input file from Fig. 6(B):**
```
MC
Output: fig6b
Particle 1: sio2.txt
Particle 2: air.txt
Matrix 1: acr.txt
Solar: AM15.txt
Photons: 100000
Start: 0.25
End: 2.5
interval: 0.005

Sim 1
Layer 1
Matrix 1
T: 50
Particle 2
D: 0.7
VF: 60

Sim 2
Layer 1
Matrix 1
T: 50
Particle 1
D: 0.9
VF: 60

Sim 3
Layer 1
Matrix 1
T: 50
Particle 1
D: 1.1
VF: 60
```